\documentclass[aps,chaos,floats,twocolumn,showpacs,superscriptaddress,reprint]{revtex4-1}

\usepackage{graphicx}
\usepackage{hyperref}
\usepackage{amsmath}
\usepackage{amssymb}
\usepackage{color}

\newcommand{\R}{\mathbb{R}}

\newcommand{\N}{\mathbb{N}}

\newcommand{\new}{\textrm{new}}

\newtheorem{theorem}[equation]{Theorem}

\begin{document}

\title{Eigenvector centrality of nodes in multiplex networks}

\author{Luis~Sol\'a}
\author{Miguel~Romance}
\author{Regino~Criado}
\author{Julio~Flores}
\author{Alejandro~Garc\'{\i}a~del~Amo}
\affiliation{Department of Applied Mathematics, Rey Juan Carlos University, Madrid (Spain)}
\affiliation{Center for Biomedical Technology (CTB), Technical University of Madrid (Spain)}
\author{Stefano~Boccaletti}
\affiliation{CNR Institute for Complex Systems, Via Madonna del Piano 10, 50019 Sesto Fiorentino, Florence (Italy)}

\date{\today}

\begin{abstract}
We extend the concept of eigenvector centrality to multiplex networks, and introduce several alternative parameters that quantify the importance of nodes in a multi-layered networked system, including the definition of vectorial-type centralities. In addition, we rigorously show that, under reasonable conditions, such centrality measures exist and are unique. Computer experiments and simulations demonstrate that the proposed measures provide substantially different results when applied to the same multiplex structure, and highlight the non-trivial relationships between the different measures of centrality introduced.
\end{abstract}

\maketitle


\textbf{Many biological, social, and technological systems find a suitable representation as complex networks, where nodes represent the system's constituents and edges account for the interactions between them \cite{Boccaletti06,newman,Newman,estrada-et-al,Wasserman}. In the general case, the nodes' interactions need a more accurate mapping than simple links, as the constituents of a system are usually simultaneously connected in multiple ways. For instance, in social networks, one can consider several  types of different actors' relationships:  friendship, vicinity, kinship, membership of the same cultural society, partnership or coworkership, etc.  In such a case, it is  useful to endow our network with a {\it multiplex network} structure. This representation reflects the interaction of nodes  through multiple layers of links, which cannot be  captured by the classical single-layer network representation.
This multiplex representation has long been considered by sociologists (multiplex tie \cite{Rapoport,Verbrugge,Granovetter}), and  although some results concerning multiplex networks' modeling and structure have been recently proposed  \cite{CFGGR,CRV,DSMZK,GRCVS,GRAF,Leicht,Buldyrev10,CFGGR,Brummitt12,Mucha}, the study of centrality parameteres in such networks has not yet been addressed satisfactorily.  The aim of this paper is to propose a definition of centrality in multiplex networks, and illustrate potential applications.}

\section{Notations}
\label{intro}

Along this paper, we consider a {\sl multiplex network} $\mathcal{G}$, made of $m\in\N$ {\sl layers} $G_1,\cdots,G_m$, such that each layer is a (directed or undirected) un-weighted network $G_k=(X,E_k)$, with $X=\{e_1,\cdots,e_n\}$ (i.e. all layers have the same $n\in\N$ nodes). The {\bf transpose} of the adjacency matrix of each layer $G_k$ is denoted by $A_k=(a_{ij}^k)\in \R^{n\times n}$, where
\[
a_{ij}^k =\left\{
\begin{tabular}{ll}
1  & \text{if $(e_j,e_i)\in E_k,$} \\
0  & \text{otherwise,}
\end{tabular}
\right.
\]
for $1\le i,j\le n$ and $1\le k\le m$. The {\sl projection network} associated to $\mathcal{G}$ is the graph $\overline{G}=(X,E)$, where
\[
E=\bigcup_{k=1}^m E_k.
\]
The {\bf transpose} of the adjacency matrix of $\overline{G}$ will be denoted by $\overline{A}=(\overline{a}_{ij})\in \R^{n\times n}$. Note that for every $1\le i,j\le n$
\[
\overline{a}_{ij}^k =\left\{
\begin{tabular}{ll}
1  & \text{if $a_{ij}^k\ne 0$ for some $1\le k\le m,$} \\
0  & \text{otherwise.}
\end{tabular}
\right.
\]

The paper is structured as follows. In the next section, we will introduce different heuristic arguments suggesting proper ways of measuring centrality in multiplex networks. Section III is devoted to establishing, under reasonable conditions, the existence and consistency of the proposed measures of centrality. In Section IV, we report some computer experiments and simulations showing how the introduced measures provide substantially different results when applied to the same multiplex networks. These results are discussed in the concluding section.


\section{Mathematical models for eigenvector centrality in connected multiplex networks}
\label{heuristics}

In the case of a multiplex network, the central question to be addressed is the following:  How can one take into account all the interactions between the different subnetworks (channels, communities, layers...) bearing in mind that not all of them have the same importance? It is essential, indeed, to remark that in order to get the centrality of a node it is necessary to take into account how the centrality (importance, influence,...) of a node is propagated within the whole network through different channels (layers) that are not necessarily additives. For instance, worldwide social networks (such as {\sf Facebook} or {\sf Twitter}) are characterized by very heterogeneous interactions, which are also typical of interactions among units in fields as diverse as  climate systems \cite{DSMZK}, game theory \cite{GRCVS, GRAF}, interacting infrastructures \cite{Leicht, Buldyrev10} and many others (\cite{CFGGR}, \cite{Brummitt12}).

With reference to the network $\mathcal{G}$, for each layer, one can consider the classical eigenvector centrality  $G_k$ as the principal eigenvector of  $A_k$ (if it exists). Specifically, the eigenvector centrality of a node $e_i$ within the layer $G_k$ would be the $i^{th}$ entry of the positive definite and normalized vector $c_k \in\R^n$  corresponding to the largest eigenvalue of the  matrix $A_k$. In a similar way, the eigenvector centrality of the projection network $\overline{G}$ will be the principal eigenvector of $\overline{A}$. The existence and uniqueness of these vectors are guaranteed by the Perron-Frobenius theorem for any symmetric matrix with positive entries.

Interestingly, the Perron-Frobenius theorem can conveniently be extended to multiplex networks, leading to even deeper concepts of nodes' centrality. We remark that other extensions of the Perron-Frobenius theorem have been proposed for hypergraphs and nonnegative tensors (\cite{MichoelNacthergaele, YangYang}).

Once all the eigenvector centralities are computed, one can consider the {\sl independent layer} eigenvector-like centrality of $\mathcal{G}$ (abbreviated as the independent layer centrality of $\mathcal{G}$) as the matrix
\[
C=
\left(
      \begin{array}{c|c|c|c}
            c_1 & c_2 & \dots & c_m
      \end{array}
\right)
\in \R^{n\times m}.
\]
Notice that $C$ is column stochastic, since $c_k>0$ and $\|c_k\|_1=1$ for every $1\le k\le m$.

Bearing in mind that the centrality (importance) of a node must be proportional to the centrality of its neighbors (lying on all layers), and considering that all layers have the same importance, one has that
$$
\forall i,j\in X,\,\,\, c(i) \varpropto c(j)\,\,\, \mbox{ if }(j\to i)\in G_\ell,\,\,\, \ell\in\{1,\dots,m\}.
$$
This allows defining the {\sl uniform eigenvector-like centrality} (abbrev. the uniform centrality)  as  the positive and normalized eigenvector $\widetilde c\in \R^n$ (if it exists) of the matrix $\widetilde A$ given by
\[
\widetilde A=\sum_{k=1}^m A_k.
\]

This situation occurs, for instance, in social networks, where different individuals may have different relationships with other people, while one is generically interested in measuring the centrality of the network of acquaintances.

Going a step further, one may consider that layers are associated with different levels of importance (or influence) in different layers of the network, and to include this sort of information in the matrix accounting for the mutual influence between layers. Thus, in order to calculate the importance (or influence) of a node within a specific layer, one must also take into account also all other layers, as some of them may be relevant for that calculation. Consider, for instance, the case of a boss going to the same gym as one of his employees: the relationship between the two fellows within the gym layer has a totally different nature from that occurring inside the office layer, but the role of the boss (i.e. his centrality) in this case can  be even bigger than if he was the only one person of the office frequenting that gym. In other words, one needs to consider the situation where the influence amongst layers is heterogeneous.

To this purpose, one can introduce an {\sl influence matrix} $W=(w_{ij})\in \R^{m\times m}$ as a non-negative matrix $W\ge 0$ such that $w_{ij}$ measures the influence {\it of} the layer $G_j$ {\it on} the layer $G_i$. Once $\mathcal{G}$ and $W=(w_{ij})$ have been fixed, one then defines the {\sl local heterogeneous eigenvector-like centrality} of  $\mathcal{G}$ (abbrev. the local heterogeneous centrality of  $\mathcal{G}$)  on each layer $G_k$ ($1\le k\le m$) as a positive and normalized eigenvector $c^{\star}_k\in \R^n$ (if it exists) of the matrix
\[
A^{\star}_k=\sum_{j=1}^m w_{kj}A_j.
\]
Once again, the {\sl local heterogeneous eigenvector-like centrality} (abbreviated as local heterogeneous centrality) matrix of the multiplex network $\mathcal{G}$ is defined as
\[
C^{\star}=
\left(
      \begin{array}{c|c|c|c}
        c^{\star}_1 & c^{\star}_2 & \dots & c^{\star}_m
      \end{array}
\right)
\in \R^{n\times m}.
\]

Another important aspect to be elucidated is that, in general, the centrality of a node $e_i$ within a specific layer $k$ may depend not only on the neighbors that are linked to $e_i$ within the layer $k$, but also to all other neighbors of $e_i$ that belong to the other layers. That is the case of scientific citations in different areas of knowledge; indeed, imagine two scientists (a chemist and a physicist) and one of them has been awarded the Nobel Prize: the importance of the other scientist will significantly increase, even though the Nobel prize laureate had few citations within the other researcher's area. This heuristic argument leads to the introduction of another concept of centrality: Given a multiplex network $\mathcal{G}$ and an influence matrix $W=(w_{ij})$, the {\sl global heterogeneous eigenvector-like centrality} of  $\mathcal{G}$ (abbrev. global centrality of $\mathcal{G}$) is defined as a positive and normalized eigenvector $c^{\otimes}\in \R^{nm}$ (if it exists) of the matrix
\[
A^{\otimes}=
\left(
\begin{array}{c|c|c|c}
  w_{11}A_1 & w_{12}A_2 & \cdots & w_{1m}A_m \\ \hline
  w_{11}A_1 & w_{22}A_2 & \cdots & w_{2m}A_m \\ \hline
  \vdots    & \vdots    & \ddots & \vdots    \\ \hline
  w_{m1}A_1 & w_{m2}A_2 & \cdots & w_{mm}A_m
\end{array}
\right) \in \R^{(nm)\times (nm)}.
\]
Note that $A^{\otimes}$ is the Khatri-Rao product of the matrices
\[
W=
\left(
  \begin{array}{c|c|c}
    w_{11} & \cdots & w_{1m} \\ \hline
    \vdots & \ddots & \vdots \\ \hline
    w_{m1} & \cdots & w_{mm}
  \end{array}
\right)
\text{ and }
\left(
\begin{array}{c|c|c|c}
A_1 & A_2 & \cdots & A_m
\end{array}
\right).
\]
In analogy with what has been one before, if one introduces the notation
\[
c^{\otimes}=
\left(
\begin{array}{c}
  c^{\otimes}_1 \\ \hline
  c^{\otimes}_2 \\ \hline
  \vdots        \\ \hline
  c^{\otimes}_m
\end{array}
\right),
\]
with $c^{\otimes}_1,\cdots, c^{\otimes}_m\in\R^n$, then one can define the {\sl global heterogeneous eigenvector-like centrality matrix} of $\mathcal{G}$ as the matrix given by
\[
C^{\otimes}=
\left(
    \begin{array}{c|c|c|c}
        c^{\otimes}_1 & c^{\otimes}_2 & \dots & c^{\otimes}_m
    \end{array}
\right)
\in \R^{n\times m}.
\]
Note that, in general $C^{\otimes}$ is neither column stochastic nor row stochastic, but the sum of all the entries of $C^{\otimes}$ is 1.

Note also that the matrix $A^\otimes$ may be interpreted as a linear operator from the tensor product $\R^n\otimes\R^m$ to itself, form which $c^\otimes$ is its normalized principal eigenvector. Using a tensor algebra approach to represent networks with different types of interactions is not new. For example, a multilinear version of Perron-Frobenius Theorem may be used to define the centrality of uniform hypergraphs (see, for instance, \cite{PearsonZhang}); furthermore, a Perron-Frobenius-type Theorem for general (not necessarily uniform), irreducible hypergraphs has been proved by \cite{MichoelNacthergaele}.


\section{Existence and consistency}
\label{existenz}

Let us now move to discussing the conditions that guarantee the existence and uniqueness of the centrality measures introduced in the previous section.

The natural question here is whether the strong connectedness of the projected graph $\overline{G}$ or, equivalently, the irreducibility of the nonnegative matrix $\overline{A}$, is a sufficient condition for the existence and uniqueness of our centralities measures. One can make use of the Perron-Frobenius theorem, as well as on irreducible matrices and strongly connected graphs, for which we refer the interested reader to Ref. \cite{Meyer}.
In fact, recalling that the graph determined by $\widetilde{A}=\sum_{k}A_k$ coincides with the projected graph of the network, in the case of the Uniform Centrality we immediately get the following

\begin{theorem}\label{thm:UCexist}
If the projected graph $\overline{G}$ of a multiplex network $\mathcal{G}$ is strongly connected, then the Uniform Centrality $\widetilde{C}$ of $\mathcal{G}$ exists and is unique.
\end{theorem}

The case of the Local Heterogeneous Centrality is similar, as every row $C^\star_\ell$ of the matrix $C^\star$ is the principal normalized eigenvector of a linear combination $A^\star_\ell=\sum_kw_{k\ell}A_k$. In particular, if $W$ is positive, the graph associated to every $A^\star_\ell$ is the projected graph of the multiplex network, hence one get also

\begin{theorem}\label{thm:LHCexist}
If the projected graph $\overline{G}$ of a multiplex network $\mathcal{G}$ is strongly connected, and $W>0$ then the Local Heterogeneous Centrality $C^\star$ of $\mathcal{G}$ exists and is unique.
\end{theorem}

A more delicate case is that of the Global Heterogeneous Centrality, that is constructed upon the principal normalized eigenvector of the matrix
$$
A^\otimes=\left(\begin{array}{c|c|c|c}
w_{11}A_1&w_{12}A_2&\cdots&w_{1m}A_m\\\hline
w_{21}A_1&w_{22}A_2&\cdots&w_{2m}A_m\\\hline
\vdots&\vdots&\ddots&\vdots\\\hline
w_{m1}A_1&w_{m2}A_2&\cdots&w_{mm}A_m
\end{array}
\right).
$$

Such a matrix is the transpose of the adjacency matrix of a graph with $nm$ nodes that we denote by $G^\otimes=(X^\otimes,E^\otimes)$, where $X=\left\{e_{ik},i=1,\dots,n,\,\,k=1,\dots,m\right\}$ and $(e_{j\ell}, e_{ik})\in E^\otimes$ iff $w_{k\ell}a^\ell_{ij}\neq 0$. Unfortunately, even if the projected graph of a multiplex network $\mathcal{G}$ is strongly connected and $W$ is positive, the graph $G^\otimes$ is not, in general, strongly connected. In fact one can easily check that this is already the case for the example in which $\mathcal{G}$ consists of two nodes and two layers, with matrices:
$$
A_1=\left(\begin{array}{cc}0&0\\0&1\end{array}\right),\quad
A_2=\left(\begin{array}{cc}0&1\\0&0\end{array}\right).
$$

Nevertheless, it is still possible to infer the existence and unicity of $C^\otimes$  from the strong-connectedness of $\overline{G}$ and the positivity of $W$. Indeed, one has  first to notice that, if $\overline{G}$ is strongly connected and $W$ is positive, then $G^\otimes$ satisfies:
$$
(e_{j\ell}, e_{ik})\in E^\otimes\iff a^\ell_{ij}\neq 0\iff (e_j, e_i)\in E_\ell.
$$
Now, we denote a node $e_{j\ell}$ of $G^\otimes$ as a $\otimes$-{\it sink} when $a_{ij}^\ell=0$ for all $i$, so that the corresponding column of $A^\otimes$ is identically zero. If a node $e_{j\ell}$ is not a $\otimes$-sink, we claim that, given any other node $e_{ik}$ there exists a path in $G^\otimes$ going from $e_{j\ell}$ to $e_{ik}$.

Assuming $\overline{G}$ to be strongly connected, there exist then indices $i_1=j,i_2,\dots,i_r=i$ such that, for every $s\in\{1,\dots,r-1\}$, there exists an index $\ell_s\in\{1,\dots,m\}$ for which $a^{\ell_s}_{i_{s+1} i_{s}}\neq 0$. Thus, by construction, $(e_{i_s\ell_s},e_{i_{s+1}\ell_{s+1}})\in E^\otimes$ for all $s$, and this finishes the proof of the latter claim.

{From} these arguments, one may easily deduce that the normal form of the matrix $A^{\otimes}$ (cf. \cite[p. 46]{Varga}) is written as
$$
N=P\cdot A^\otimes\cdot P^t=\left(\begin{array}{ccc|ccc}
0&\cdots&0&\star&\cdots&\star\\
\vdots&\ddots&\vdots&\vdots&&\vdots\\
0&\cdots&0&\star&\cdots&\star\\\hline
0&\cdots&0&&&\\
\vdots&&\vdots&&\Huge{B}&\\
0&\cdots&0&&&
\end{array}\right),
$$
where $P$ is a permutation matrix and $B$ is an irreducible nonnegative matrix, to which the Perron-Frobenius Theorem can be applied. It follows that the spectrum of $A^\otimes$ is the union of the spectrum of $B$ and $\{0\}$, and that $A^\otimes$ has a unique normalized  eigenvector associated to $\rho(A^\otimes)=\rho(B)$. Summing up, we get the following

\begin{theorem}\label{thm:GHCexist}
If the projected graph $\overline{G}$ of a multiplex network $\mathcal{G}$ is strongly connected, and $W>0$ then the Global Heterogeneous Centrality $C^\otimes$ of $\mathcal{G}$ exists and is unique.
\end{theorem}

We now discuss the consistency of our definitions in a variety of special cases.

\noindent{\bf Monoplex networks.} It is  straightforward to demonstrate that on a monoplex network (i.e. a multiplex network consisting of only one layer) our three concepts of multiplex centrality coincide with the usual eigenvector centrality of the layer.

\noindent{\bf Identical layers.} Let $\mathcal{G}$ be a multiplex network for which $A_k=A_\ell$ for every $1\le k,\ell\le m$, and note that $A_k=\overline{A}$, for every $k$, so that the Uniform Centrality of $\mathcal{G}$ coincides with the Eigenvector Centrality of every layer $G_k$. Assuming that every row of $W$ is nonnegative (in particular if $W>0$) it is also clear that every column of the Local Heterogeneous Centrality  $C^\star$ coincides with the Uniform Centrality $\overline{C}$ of $\mathcal{G}$.

The case of the Global Heterogeneous Centrality is slightly different. If all the layers are identical, the matrix $A^\otimes$ coincides with the so called {\it Kronecker product} of the matrices $W$ and $\overline{A}$. It is well known (see for instance \cite[Ch.~2]{Steeb}) that the spectral radius of $A^\otimes$ is then equal to $\rho(W)\rho(\overline{A})$ and that its normalized principal eigenvector is the Kronecker product of the normalized principal eigenvectors $C_W$ of $W$ and $\overline{C}$ of $\overline{A}$. In terms of matrices, this is equivalent to say that $C^\otimes=C_W^t\cdot \overline{C}$. In particular, the normalization of all the columns of $C^\otimes$ equals $\overline{C}$.

\noindent{\bf Starred layers.} We  finally consider the case in which the multiplex network $\mathcal{G}$ contains exactly $m=n$ layers, satisfying that the layer $G_k$ consists of a set of edges coming out of the node $e_k$. In other words, $a^k_{ij}=0$ if $j\neq k$.
In this case there exists a permutation matrix $P$ such that:
$$
P\cdot A^\otimes\cdot P^t=\left(\begin{array}{ccc|ccc}
0&\cdots&0&\star&\cdots&\star\\
\vdots&\ddots&\vdots&\vdots&&\vdots\\
0&\cdots&0&\star&\cdots&\star\\\hline
0&\cdots&0&&&\\
\vdots&&\vdots&&\hspace{-0.3cm}{W\circ \overline{A}}\hspace{-0.3cm}&\\
0&\cdots&0&&&
\end{array}\right),
$$
where $W\circ \overline{A}$ is the Hadamard product (see, for example \cite{HJ}) of  $W$ and $\overline{A}$ (i.e. $(W\circ \overline{A})_{ij}=w_{ij}\overline{a}_{ij}$). In particular the Global Heterogeneous Centrality of $\mathcal{G}$ is the diagonal $n\times n$ matrix whose diagonal is the eigenvector centrality of $W\circ \overline{A}$. Note that $W\circ \overline{A}$ can be interpreted as the transpose of the matrix of the graph $\overline{G}$, in which the edge going from $e_j$ to $e_i$ has been assigned a weight equal to $w_{ij}$. In this sense the eigenvector centrality of a weighted graph can be seen as a particular case of the Global Heterogeneous Centrality.

\section{Comparing centralities of a multiplex network}
\label{comparing}


In the following two sections we will compute and compare the different types of centrality measures that we have defined for some examples, constructed upon both real and synthetic data. We will start by describing two ways of comparing centrality measures, and then we will apply them to a real example of social multiplex network.

If we take a network of $n$ nodes $\{e_1,\cdots,e_n\}$ and  consider two centrality measures $c,c'\in\R^n$ such that the $i$-th coordinate of $c$ and $c'$ measure the centrality of node $v_i$ for every $1\le i\le n$, one way of measuring the correlation between $c$ and $c'$ is by computing $\|c-c'\|$ for some norm $\|\cdot\|$. While $\|c-c'\|$ measures the discrepancy between $c$ and $c'$, its value is not representative of the {\it real} information about the correlation between $c$ and $c'$. Note, indeed, that one of the main features of the centrality measures is the fact that they produce {\sl rankings}, i.e. in many cases the crucial information obtained from a centrality measure is the fact that a node $v_i$ is more relevant than another node $v_j$, and this ordering is more important than the actual difference between the corresponding centrality of nodes $v_i$ and $v_j$. Hence, if we want to analyze the correlations among a set of centrality measures, we should study in detail the correlations between the associated rankings.

The literature suggests various alternative ways to study the correlations between two rankings $r$ and $r'$, two standard ones being the Spearman's rank correlation coefficient $\rho(r,r')$ and the Kendall's rank correlation coefficient $\tau(r,r')$. If we consider two centrality measures $c,c'\in\R^n$ of a network with nodes $\{e_1,\cdots,e_n\}$, then each centrality measure $c$ and $c'$ produces a ranking of the nodes that will be denoted by $r$ and $r'$ respectively. The Spearman's rank correlation coefficient \cite{Spearman} between two centrality measures $c$ and $c'$ is defined as
\[
\rho(c,c')=\rho(r,r')
=\frac{\sum_{i=1}^n(r(v_i)-\overline{r})(r'(v_i)-\overline{r'})}{\sqrt{\sum_{i=1}^n(r(v_i)-\overline{r})^2(r'(v_i)-\overline{r'})^2}},
\]
where $r(v_i)$ and $r'(v_i)$ are the ranking of node $v_i$ with respect to the centrality measures $c$ and $c'$ respectively, $\overline{r}=\frac 1n\sum_ir(v_i)$ and $\overline{r'}=\frac1n\sum_ir'(v_i)$. Similarly, the Kendall's rank correlation coefficient \cite{Kendall}  between two centrality measures $c$ and $c'$ is defined as
 \[
 \tau(c,c')=\tau(r,r')=\frac {\tilde K(r,r')- K(r,r')}{\binom n 2},
 \]
where $ \tilde K(r,r')$ is the number of pairs of nodes $\{v_i,v_j\}$ such that they appear in the same ordering in $r$ and $r'$ and $K(r,r')$ is the number of pairs of nodes $\{v_i,v_j\}$ such that they appear in different order in rankings $r$ and $r'$. Note that both $\rho(c,c')$ and $\tau(c,c')$ give values in $[-1,1]$. The closer $\rho(c,c')$ is to 1 the more correlated $c$ and $c'$ are, while the closer $\rho(c,c')$ is to 0 the more independent $c$ and $c'$ are (and similarly for $\tau(c,c')$). In addition, if $\rho(c,c')$ (or $\tau(c,c')$) is close to $-1$ then $c$ and $c'$ are anti-correlated.

A further remark comes from the fact that the centrality measures introduced so far are very different from one another, and therefore one has to carefully describe how to compare them. Indeed, on one hand, some {\sl scalar} measures introduced in section~\ref{heuristics} (the centrality of the node in the network) associate a single number to each node of the network, while on the other hand, other {\sl vectorial} measures assign a vector to each node $v_i$ (with each coordinate of the vector measuring the centrality of the node $v_i$ as an actor of a different layer of the multiplex network).  Actually, for a multiplex network $\mathcal{G}$ of $n$ nodes, two scalar centralities (the eigenvector centrality $\overline{c}\in\R^n$ of the projection graph, and the uniform eigenvector-like centrality $\widetilde c\in\R^n$) and  three vectorial centralities (the independent layer centrality $C\in\R^{n\times m}$, the local heterogeneous centrality $C^{\star}\in\R^{n\times m}$, and the global heterogeneous centrality $C^{\otimes}\in\R^{n\times m}$)  have been proposed. To compare these different measures, the information contained in each vectorial-type centrality must be aggregated to associate a number to each node.

There are several alternative  methods for aggregating information, but we use the convex combination technique as main criterion. For a multiplex network $\mathcal{G}$ of $n$ nodes and $m$ layers, we can fix some $\lambda_1,\cdots,\lambda_m\in[0,1]$ such that $\lambda_1+\cdots+\lambda_m=1$ and  compute the aggregated scalar centralities
\[
\begin{split}
 c=&c(\lambda_1,\cdots,\lambda_m)=\sum_{j=1}^m\lambda_jc_j,\\
 c^{\star}=&c^{\star}(\lambda_1,\cdots,\lambda_m)=\sum_{j=1}^m\lambda_jc^{\star}_j,
 \end{split}
\]
where $c_j$ is the $j$th-column of the independent layer centrality $C$ and $c^{\star}_j$ is the $j$th-column of the local heterogeneous centrality $C^{\star}$. Note that the value of each $\lambda_j$ can be understood as the {\sl relative influence} of the layer $G_j$ in the aggregated scalar centrality of the multiplex network.  In our numerics, the specific value $\lambda_1=\cdots=\lambda_m=\frac 1m$ has been chosen, as we suppose that no extra information about the relative relevance of each layer is available, and therefore the influence of each of them is considered equivalent. Note that $c$ and $c^{\star}$ are normalized, since $C$ and $C^{\star}$ are column-stochastic.

The case of the global heterogeneous centrality is different, since $C^{\otimes}$ is not column-stochastic. In this case, since the sum of all entries of $C^{\star}$ is 1, it is enough to take
\[
c^{\otimes}=\sum_{j=1}^mc^{\otimes}_j,
\]
where $c^{\otimes}_j$ is the $j$th-column of the global heterogeneous centrality $C^{\otimes}$. Consequently, the {\sl relative influence} of each layer $G_j$ can be defined as $\|c^{\otimes}_j\|_1$ (i.e. the sum of all the coordinates of $c^{\otimes}_j$).

Once all the vectorial measures have been aggregated (and the setting unified), we discuss the ranking comparisons. In addition to the actual correlation among the centrality measures, we analyze  the influence of the matrix $W$ (called {\sl influence matrix} in section~\ref{heuristics}) used in the definition of the  local heterogeneous centrality $C^{\star}$ and in the global heterogeneous centrality $C^{\otimes}$. Since this matrix $W\in\R^{m\times m}$ is non-negative we consider two families of matrices $\{W_1(q)\}$ and $\{W_2(q)\}$ given for every $0\le q\le 1$ by
\[
W_1(q)=
       \left(
       \begin{array}{cccc}
        1      & q      & \cdots & q \\
        q      & 1      & \cdots & q \\
        \vdots & \vdots & \ddots & \vdots\\
        q     & q     & \cdots & 1
       \end{array}
       \right),\hfill
W_2(q)=
       \left(
       \begin{array}{cccc}
        1      & q      & \cdots & q\\
        q^2    & 1      & \cdots & q\\
        \vdots & \vdots & \ddots & \vdots\\
        q^2    & q^2    & \cdots & 1
       \end{array}
       \right).
\]
Note that while each $W_1(q)$ corresponds to a symmetric influence among the layers, each $W_2(q)$ models an asymmetric influence among the layers of the multiplex network.

\begin{figure*}[t]
$\,$\hfill
\includegraphics[width=0.4\textwidth]{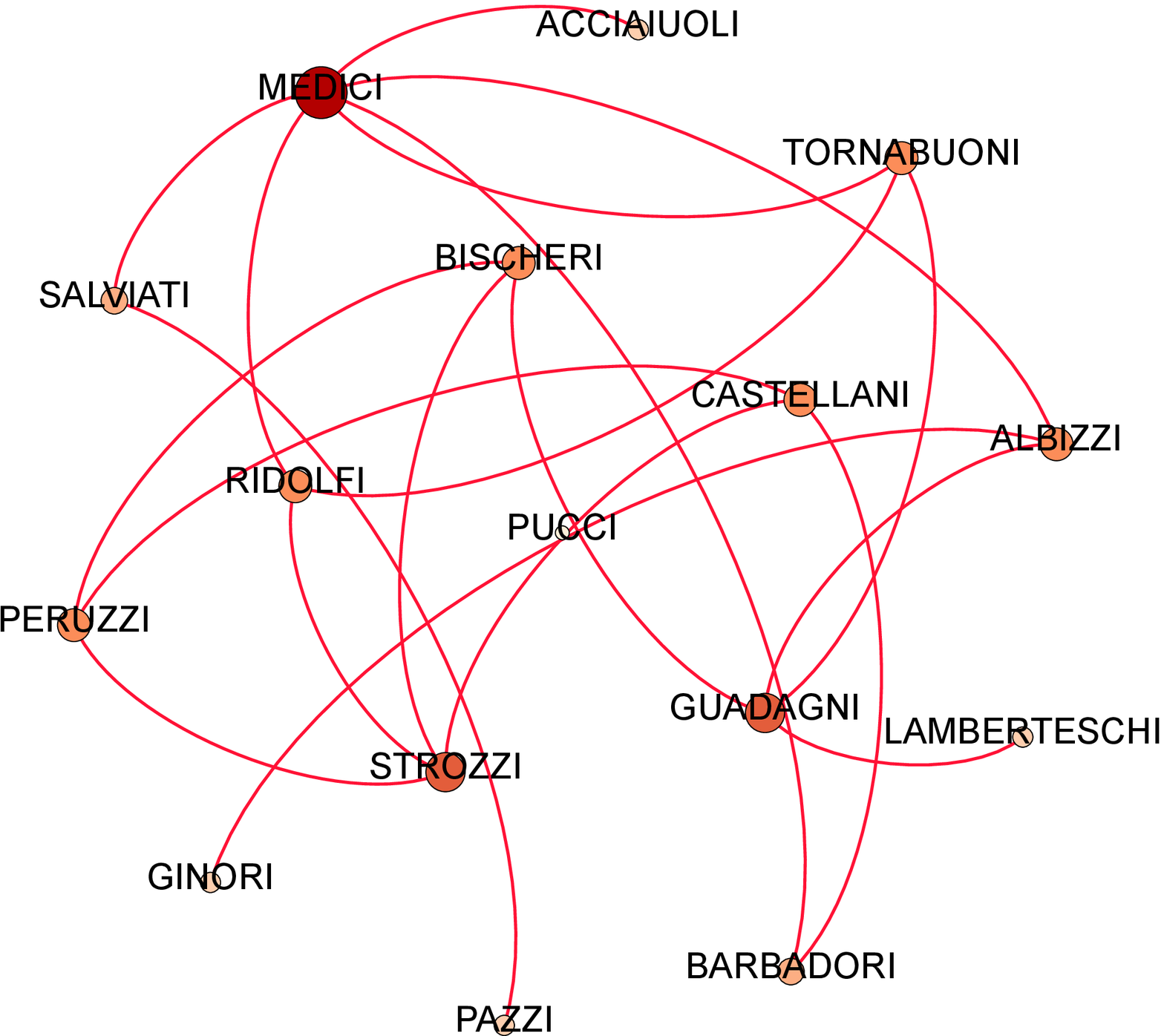}\hfill
\includegraphics[width=0.4\textwidth]{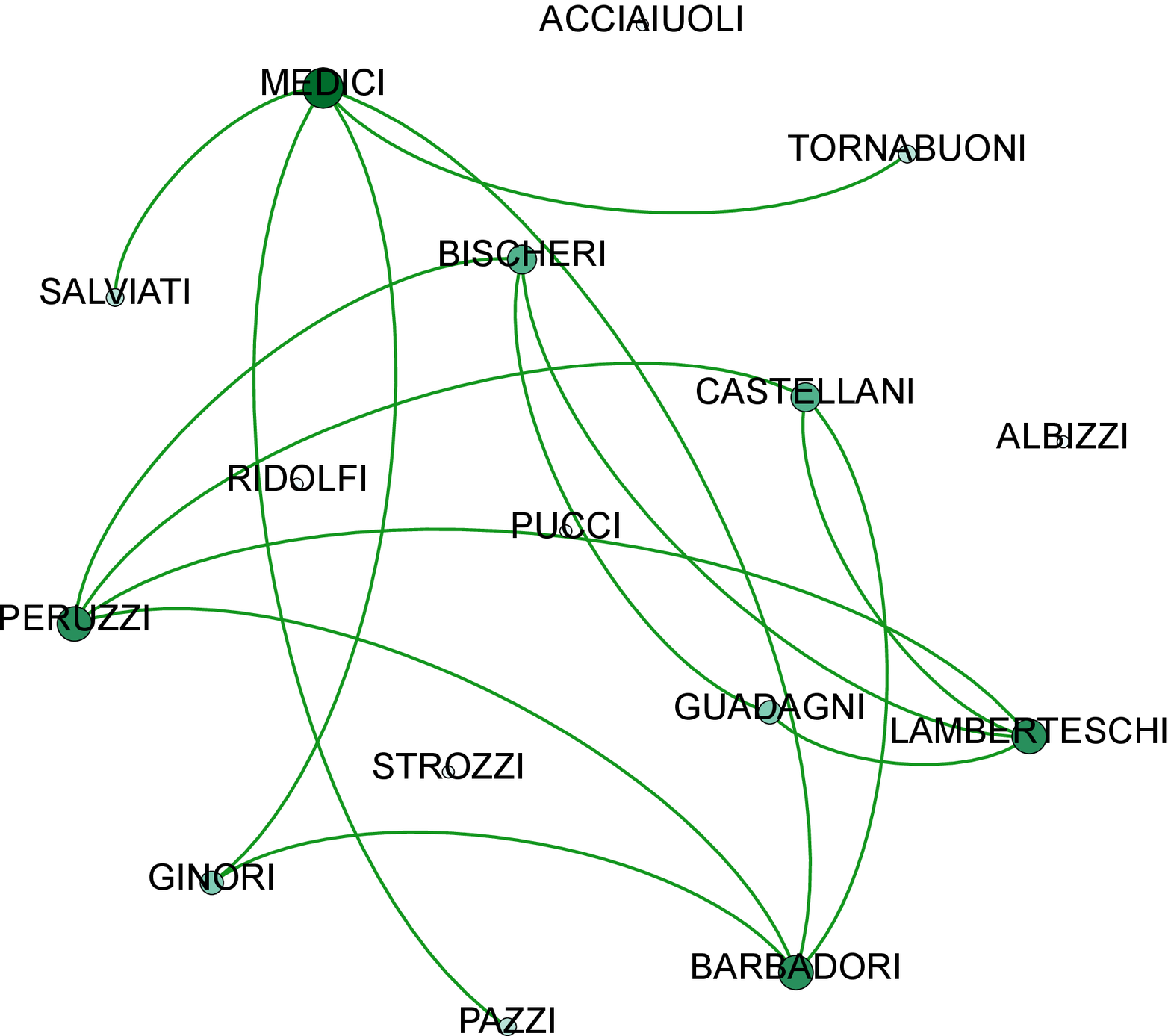}\hfill$\,$\hfill
\caption{\label{Florence01}
The business layer (on the left) and the marriage layer (on the right) of the social multiplex network of the Renaissance Florentine families.}
\end{figure*}

We apply now our methods of comparison of the different centralities to a classic example: the social network of the Renaissance Florentine families in $1282-1500$. The dataset of the network (that are available in \cite{ucinet}) collects information about marriage and business ties among sixteen Renaissance Florentine families. This social system can be modelled as a multiplex network with two layers: one related with the business ties (specifically, recorded financial ties such as loans, credits and joint partnerships) and other that shows the marriage ties in the total dataset of sixteen families (see \cite{BrePa, Padgett}). These two layers are represented in Figure~\ref{Florence01}.

\begin{figure*}[t]
\includegraphics[width=\textwidth]{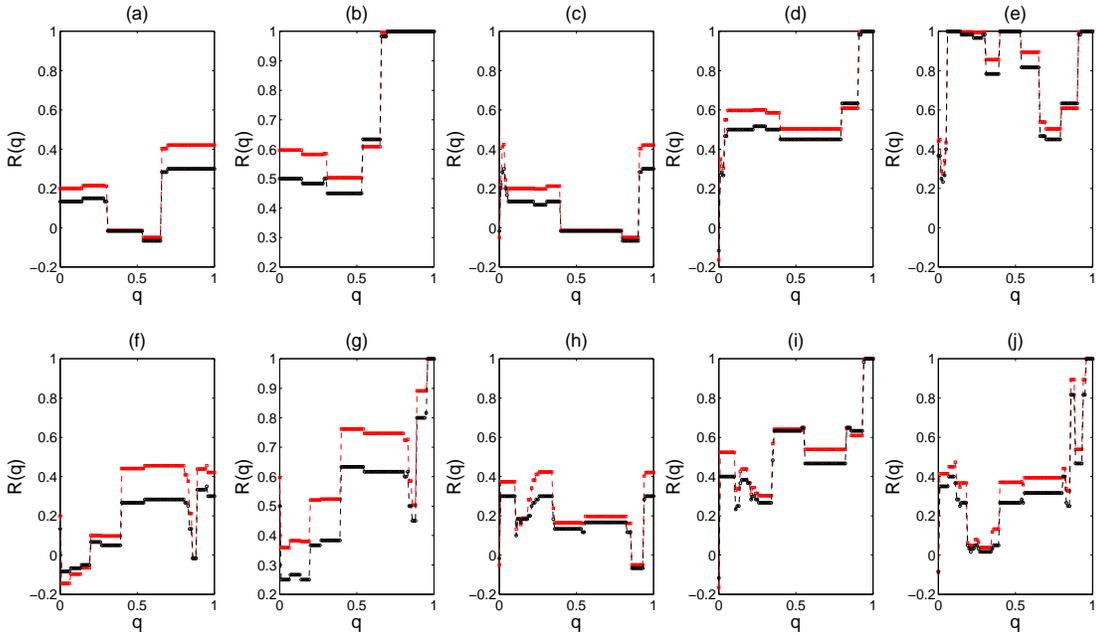}
\caption{\label{Florence_centrality}
Ranking comparisons for the eigenvector centrality measures for the social multiplex network of the Renaissance Florentine families with the family of {\it symmetric} influence matrices of type $W_1(q)$ (panels from (a) to (e)) and with the family of {\it non-symmetric} influence matrices of type $W_2(q)$ (panels from (f) to (j)). Panels in the first and second column show the ($q$-dependent) correlations between the eigenvector centrality of the projection and the uniform centrality vs. the local heterogeneous centrality, respectively. Panels in the third and fourth column show the ($q$-dependent) correlations between the eigenvector centrality of the projection and the uniform centrality vs. the global heterogeneous centrality, respectively. Finally, the fifth column shows the correlation between the local and the global heterogeneous centrality. In all  panels, Spearman and Kendall coefficient are respectively depicted in red and black.}
\end{figure*}

The comparisons among the different centrality measures for the social multiplex network of the Renaissance Florentine families is presented in Figure~\ref{Florence_centrality}. More precisely, we represent the $q$-dependent Spearman (in red) and Kendall (in black) correlation coefficients among the eigenvector centrality of the projection graph, the uniform centrality, the local heterogeneous centrality and the global heterogeneous centrality, in this particular example. 

\section{Numerical testings}
\label{numerical}

In this section we illustrate the different behaviour of the introduced centrality measures by testing them against a class of randomly generated multiplex networks. To do so, instead of considering particular taylor-made examples, we consider random networks from a class of scale-free assortative-inspired synthetic graphs (cf. \cite{CFGGR}), that we will describe later on.

First of all, we briefly describe the method used to construct the synthetic multiplex networks used in the numerical testing, which corresponds to the {\sl model II} of Ref. \cite{CFGGR}. The model is inspired by the Barab\'asi-Albert preferential attachment model \cite{Albert1} as well as by several bipartite networks models such as the collaboration network model proposed by J.J.\,Ramasco {\em et al.} \cite{ramasco}, or the sublinear preferential attachment bipartite model introduced by M.\,Peltom\"{a}ki and M.\,Alava \cite{peltomaki}. It consists of a growing random model determined by the following rules:
\begin{itemize}
 \item[{\it (i)}] {\em Model parameters}. The model has three main parameters: $n$, $m$ and $p_{\new}$. We set $n\in \N$  as the minimal number of nodes in the multiplex network and $2\le m\le n$ as the number of {\sl active nodes} in each layer ({\em i.e.} nodes that will produce links in each layers). Note that if we take $m=2$, we recover the Barab\'asi-Albert model \cite{Albert1}. In this model $m$ will be fixed, but the results are similar for other non-negative integer random variable. Finally, we set $p_{\new}\in(0,1]$ as the probability of joining a new node to the growing multiplex network during its construction.

\item[{\it (ii)}] {\em Initial conditions}. We start with a seed multiplex network made of a single layer $G_0$ of $m$ nodes that are linked all to all, ({\em i.e.} $G_0$ is the complete graph $K_m$). We can replace the {\sl all-to-all} structure by any other structure (such as a scale free or a Erd\H{o}s-R\'enyi network), but the results obtained are similar. This initial layer $G_0$ will be removed from the final multiplex network $\mathcal{G}$, since the {\sl all-to-all} structure would make the eigenvector-centrality of the projection graph a bisector.

 \item[{\it (iii)}] {\em Layer composition}. At each time step $t$, a new layer $G_t$ of $m$ nodes is added to the multiplex network.
We start by randomly choosing an existing node of the multiplex network with a probability proportional to its degree ({\em preferential election}) that we call the {\sl coordinator node}. Therefore if at step $t-1$, the set of nodes of the multiplex network is $\{v_1,\ldots,v_n\}$, and $k_i$ denotes the degree of node $v_i$ at time $t-1$ in the projection network, then we choose the node $v_i$ randomly and independently with probability
     \[
     p_i=\frac {k_i}{\sum_{j=1}^{n} k_j}.
     \]
Once the coordinator node has been chosen, each of the remaining $m-1$ active nodes of $G_t$ will be a new node with probability $p_{\new}$ and an existing node with probability $(1-p_{\new})$. Already existing nodes are added by choosing them uniformly and independently. Note that we can replace the uniform random selection by other random procedures (such as preferential selection), but the random tests done suggest that the multiplex network obtained have statistically the same structural properties when $n$ is large enough (see \cite{CFGGR}). At this step, we have chosen $m$ nodes $\tilde v_1, \ldots, \tilde v_m$ that will be the {\sl active nodes} of the new layer $G_t$ (i.e. nodes that will produce links in this layers).

 \item[{\it (iv)}] {\em Layer inner-structure}. After fixing the active nodes $\tilde v_1, \ldots, \tilde v_m$ of the new layer $G_t$, we have to give its links. First, we link all the active nodes to the coordinator in order to ensure that all the eigenvector-like centrality are well defined. We set new links between each pair of active nodes $v_i$ and $v_j$ (with $1<i \ne j\leq m$) by using a {\sl random assortative linking strategy} (this corresponds to the {\sl Model II in \cite{CFGGR}}). For every $2\le i\ne j\le m$, we add randomly the link $\{\tilde v_i,\tilde v_j\}$ in proportion to the number of common layers that hold simultaneously $\tilde v_i$ and $\tilde v_j$. Hence if we denote by $Q_{ij}$ the number of layers that hold simultaneously $\tilde v_i$ and $\tilde v_j$ at time step $t$ (including $G_t$) and by $q_i$ the number of layers that hold $\tilde v_i$ at time step $t$ (also including $G_t$), thus the probability of linking node $\tilde v_i$ with node $\tilde v_i$ is given by
     \[
     p_{ij}=\frac {2Q_{ij}}{q_i+q_j},
     \]
     for every $2\le i\ne j\le m$. The heuristic behind this model comes from social networks, since the relationships in a new social group are correlated with the previous relationships between the actors in other social groups \cite{Wasserman}. Hence, if two actors that belong to the new social group coincide in many (previous) groups, then the probability of being connected in this new  group is large. The model also reflects the fact that if two new actors join their first group, the probability of establishing a relationship between them is high. At the end of this step, the new layer $G_t$ is completely defined.

 \item[{\it (v)}] Finally, we repeat steps {\it (iii)} and {\it (iv)} until the number of nodes of the multiplex network is at least $n$.
\end{itemize}

After fixing all the settings of the numerical testings, we perform the comparison for three multiplex networks $\mathcal{G}_1$, $\mathcal{G}_2$ and $\mathcal{G}_3$ (constructed as above), where:
\begin{itemize}
 \item[{\it(i)}] $\mathcal{G}_1$ is a network of 102 nodes (computed with $n=100$ as initial parameter) and 13 layers of 10 nodes each ($k=10$ as initial parameter). The probability $p_{\new}=0.8$ of adding new active nodes to each layer. This is an example of a network with a relative small number of active nodes in each layer and such as each node is active in a few number of layers (since $p_{\new}=0.8$).
 \item[{\it(ii)}] $\mathcal{G}_2$ is a  network of 108 nodes (computed with $n=100$ as initial parameter) and 4 layers of 40 nodes each ($k=40$ as initial parameter). The probability $p_{\new}=0.5$ of adding new active nodes to each layer. In this case, this is a  network with a relative big number of active nodes in each layer and a balanced number of newcomers and experienced nodes as actives nodes in each layer ($p_{\new}=0.5$).
 \item[{\it(iii)}] $\mathcal{G}_3$ is a  network of 102 nodes (computed with $n=100$ as initial parameter) and 6 layers of 60 nodes each ($k=60$ as initial parameter). The probability $p_{\new}=0.1$ of adding new active nodes to each layer. In this case, this is a network with a big number of active nodes in each layer and a very low number of newcomers in each layer ($p_{\new}=0.1$).

\end{itemize}

\begin{figure*}[t]
\includegraphics[width=\textwidth]{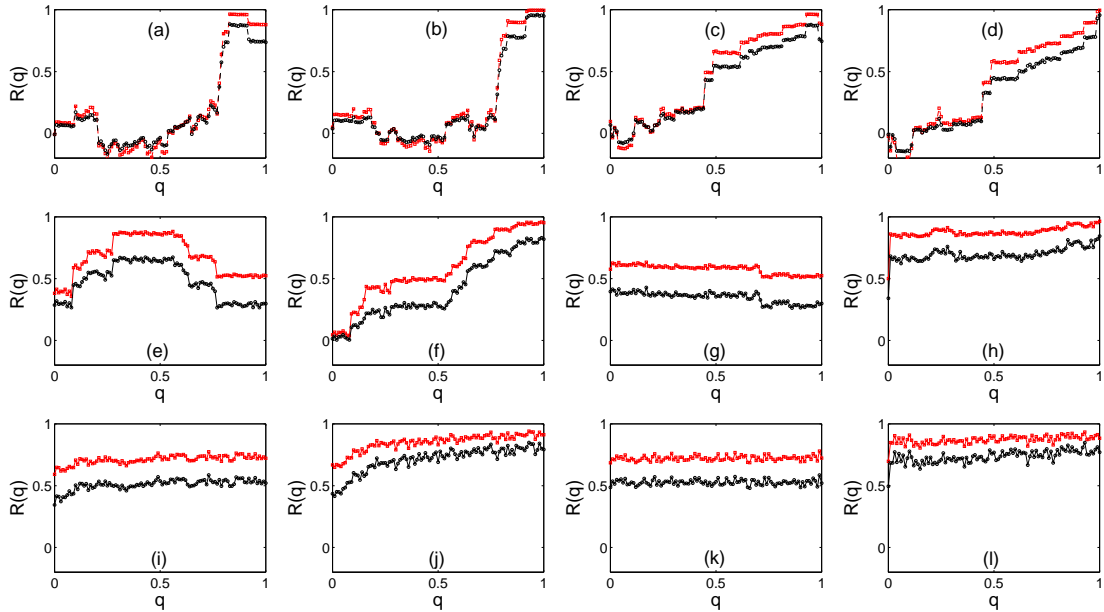}
\caption{\label{fig1}
Ranking comparison for the eigenvector centrality measures for two multiplex networks with the family of {\it symmetric} influence matrices of type $W_1(q)$. Panels (a,b,c,d), (e,f,g,h), and (i,j,k,l) respectively correspond to network $\mathcal{G}_1$, $\mathcal{G}_2$ and $\mathcal{G}_3$
(see text for details on the network construction).
Panels (a) and (b) ((c) and (d)) show the ($q$-dependent) correlations between the eigenvector centrality of the projection graph and the uniform centrality vs. the local (global) heterogeneous centrality of $\mathcal{G}_1$, respectively. Similarly, panels (e) to (h) give the same information for $\mathcal{G}_2$ and panels (i) to (l) correspond to $\mathcal{G}_3$ respectively.} In all panels Spearman and Kendall coefficient are respectively depicted in red and black.
\end{figure*}

\begin{figure*}[t]
\includegraphics[width=\textwidth]{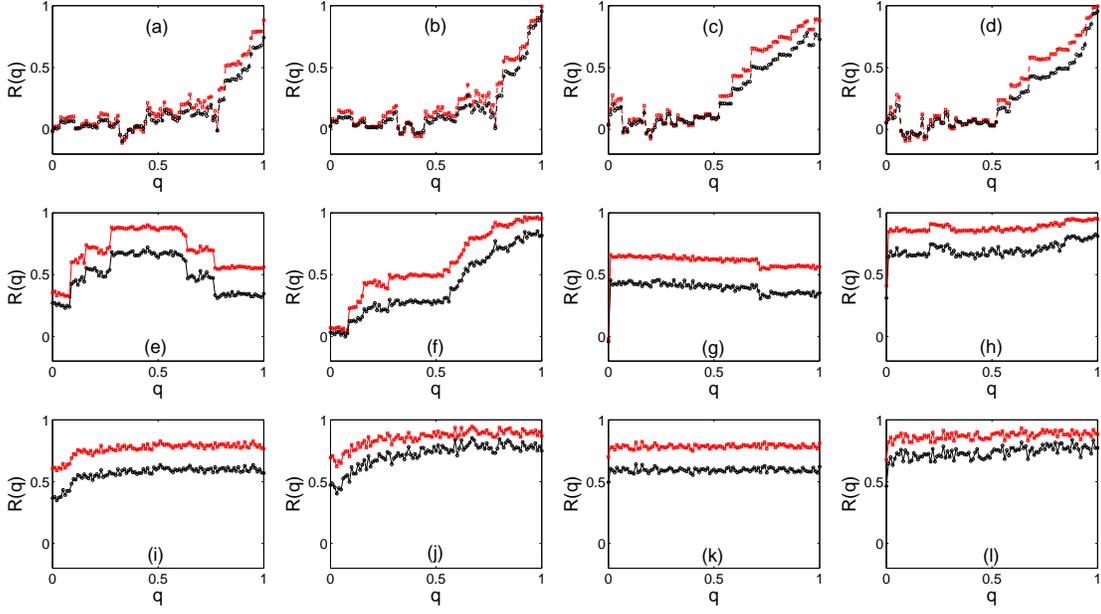}
\caption{\label{fig2}
Ranking comparison for the eigenvector centrality measures for two multiplex networks with the family of {\it non-symmetric} influence matrices of type $W_2(q)$.
Panels (a,b,c,d), (e,f,g,h), and (i,j,k,l) respectively correspond to network $\mathcal{G}_1$, $\mathcal{G}_2$ and $\mathcal{G}_3$. Panels (a) and (b) ((c) and (d)) show the ($q$-dependent) correlations between the eigenvector centrality of the projection graph and the uniform centrality vs. the local (global) panels (e) to (h) give the same information for $\mathcal{G}_2$ and panels (i) to (l) correspond to $\mathcal{G}_3$ respectively. Same stipulations as in the caption of Figure 1.}
\end{figure*}

\begin{figure*}[t]
\includegraphics[width=0.9\textwidth]{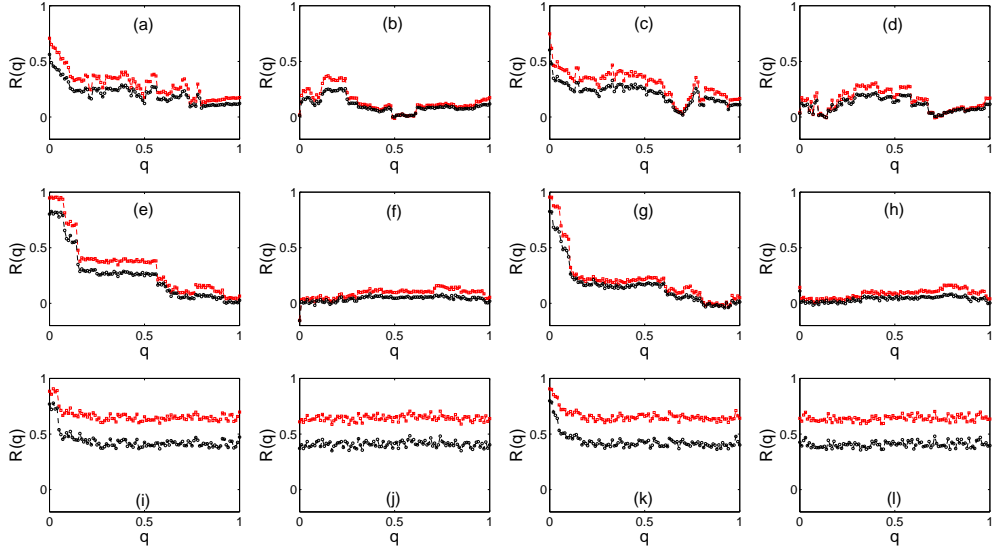}
\caption{\label{fig3}
Ranking comparison for the independent layer centrality vs. local and global heterogeneous centralities. Panels (a,b,c,d), (e,f,g,h), and (i,j,k,l) respectively correspond to network $\mathcal{G}_1$, $\mathcal{G}_2$ and $\mathcal{G}_3$. The first two columns of panels on the left correspond to the symmetric family of influence matrices $W_1(q)$ while the two on the right are for the asymmetric family of influence matrices $W_2(q)$ ($0\le q\le 1$). The first and the third columns of panels on the left show the correlations between the independent layer centrality and the local heterogeneous centrality, while  the second and the forth columns of panels on the left show the correlations between the independent layer centrality and the global heterogeneous centrality.  The Spearman coefficient is in red, and the Kendall coefficient is in black.}
\end{figure*}

\begin{figure*}[t]
\includegraphics[width=0.9\textwidth]{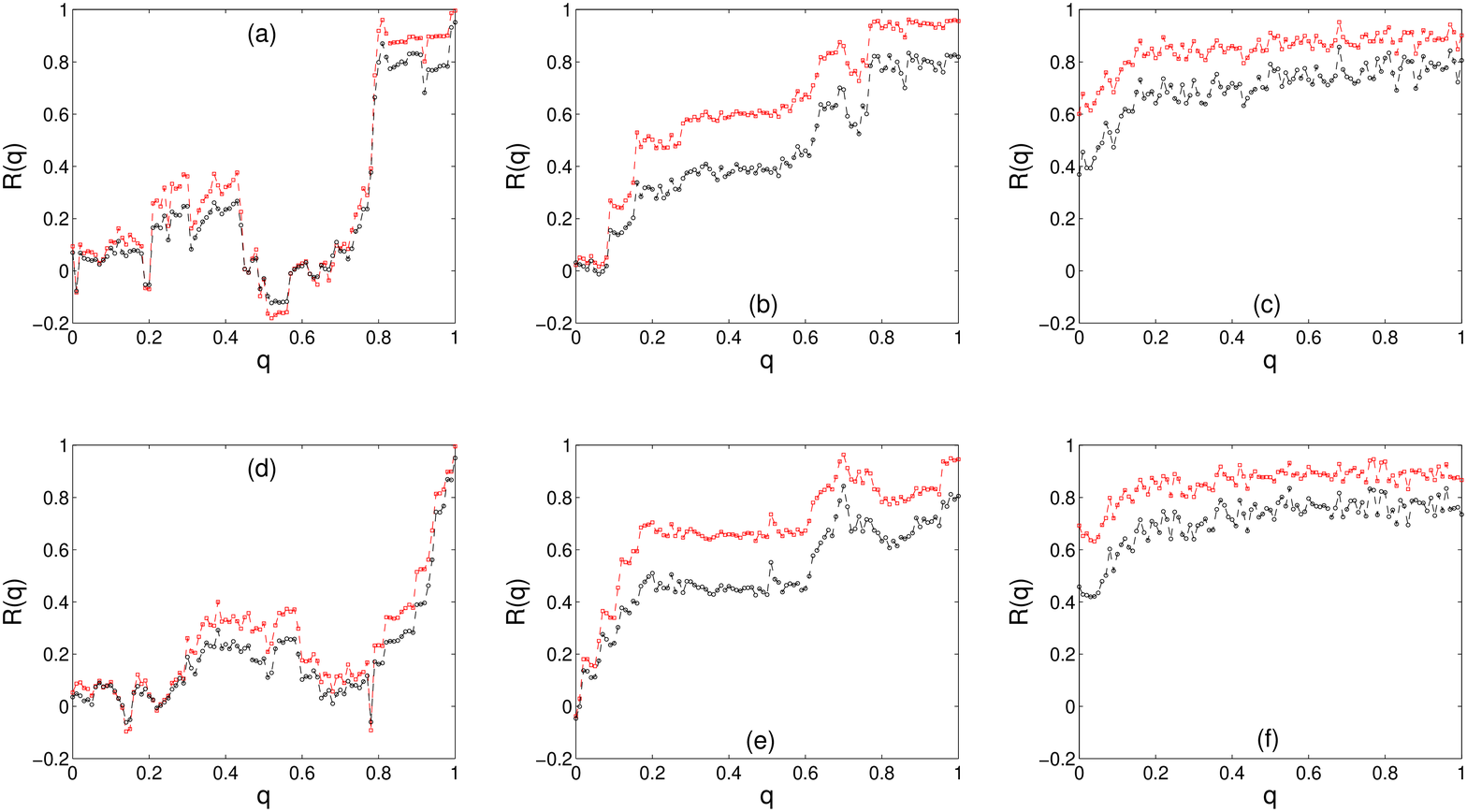}
\caption{\label{fig4}
Ranking comparison between the local heterogeneous centrality and the the global heterogeneous centrality for $\mathcal{G}_1$ (panels (a) and (d)),  for $\mathcal{G}_2$  (panels (b) and (e))and for $\mathcal{G}_3$  (panels (c) and (f)). The computation has been done with the family of \textit{symmetric} influence matrices of type $W_1(q)$ (top panels) and with the family of \textit{non-symmetric} influence matrices of type $W_2(q)$ (bottom panels). Once again, the Spearman coefficient is in red, and the Kendall coefficient is in black.}
\end{figure*}

For each of these networks we compute the correlation between the eigenvector centrality of the projection graph, and the uniform centrality vs. the local and global heterogeneous centralities. Figures (\ref{fig1}) and (\ref{fig2}) plot the dependency of these correlations with respect to the influence strength $q\in[0,1]$ in a family of \textbf{symmetric} influence matrices $W_1(q)$ (Figure~\ref{fig1}) and with respect to the influence strength $q\in[0,1]$ in a family of \textbf{non-symmetric} influence matrices $W_2(q)$ (Figure~\ref{fig2}), exhibiting a similar pattern. Note that this phenomena does not occur in the case of the example considered in section \ref{comparing}, since in this case there were deep differences between the symmetric case and the non symmetric one (see Figure \ref{Florence_centrality}). Similar results for the correlations between the heterogeneous centralities and the independent layer centrality are displayed in Figure~\ref{fig3}. Finally, we also report the local heterogeneous centrality vs. the global heterogeneous centrality, under the action of the two families of influence matrices $W_1(q)$ and $W_2(q)$ (see Figure (\ref{fig4})).

\section{Discussion and Conclusions}
\label{conclusions}

Introducing a layer structure on a complex network or, equivalently, distinguishing different types of interactions between its nodes, may significantly vary the behaviour of the network (cf. \cite{Leicht, Buldyrev10,Brummitt12}). The main goal of this paper is analysing the influence of the layer structure in some eigenvector-like centralities of multiplex networks. In order to that, we have introduced several eigenvector centralities that take into account the layer structure by means of a directed graph of influences among layers. The examples presented in the paper show that the centrality measures introduced are qualitatively different and, in particular, different from the eigenvector centrality of the projected network. In order to measure conveniently these differences, we have introduced an algorithm that produces randomly generated multiplex networks and measured the pairwise correlations of the different centralities studied, under different types of influence between layers, according to a parameter $q\in[0,1]$ and two distinct types of influence matrix. We have selected three representative examples from the family of synthetic networks analysed, and presented them here since all the numerical simulations we have performed show similar behaviour. For the multiplex examples considered, this behaviour may be described as follows:

\begin{itemize}
 \item The rankings given by the different eigenvector centrality measures introduced in the paper are qualitatively different and hence the corresponding centrality measures are also different.
 
  \item The correlations between these new eigenvector centrality measures strongly depend on the structure of the multiplex networks, including the number of layers and the number of nodes per layer.

 \item The results obtained with Spearman's and Kendall's coefficients are qualitatively equivalent in all the examples considered, although Spearman's rank is always slightly higher.
 
 \item The differences between the heterogeneous (global, local) and the flat centralities (centrality of the projected network, uniform centrality) are significantly broader for lower values of $q$.  In fact, there is a non-linear relationship between the centrality measures and the strength $q$ of the influence between layers. On the other hand,for high values of $q$, the behaviour of these particular multiplex networks is similar to the corresponding, monoplex, projected networks. In other words $q$, thought of as a measure of the multiplexity of the network, is detected by heterogeneous centrality measures.

 \item In the synthetic examples considered, the total variation with respect to $q$ of the correlation between a heterogeneous and a flat measure grows with the ratio between number of layers and number of nodes of each layer.


 \item The symmetry of the influence between layers does not play a critical role in the correlations among centrality measures in the randomly generated networks considered. However, in the example of the Florentine families (in which the number of nodes and layers is small) the differences between the symmetric and non symmetric case is significant.

\end{itemize}

In summary, we introduced several definitions of centrality measures for multiplex networks, and proved that, under reasonable conditions, these centrality measures exist and are unique (theorems \ref{thm:UCexist}, \ref{thm:LHCexist}, and \ref{thm:GHCexist}). Computer experiments and simulations performed by using the model introduced in \cite{CFGGR} show that our measures provide substantially different results when applied to the same multiplex networks. This is in agreement with the fact that each of these measures arises from a different heuristic. In this sense, the concept of multiplex network may be used to model complex networks of different kinds, so that the most appropriate kind of centrality measure shall be carefully determined in each case.

\section*{Acknowledgements}

The authors would like to thank an anonymous referee for useful suggestions, that have helped us to improve the final version of this paper. We also thank David Papo for his valuable comments. This work has been partially supported by the Spanish DGICYT under projects MTM2009-13848 and MTM2012-32670.



\end{document}